\documentclass[fleqn,usenatbib]{mnras}

\usepackage{newtxtext,newtxmath}

\usepackage[T1]{fontenc}

\DeclareRobustCommand{\VAN}[3]{#2}
\let\VANthebibliography\thebibliography
\def\thebibliography{\DeclareRobustCommand{\VAN}[3]{##3}\VANthebibliography}

\usepackage{graphicx}	
\usepackage{amsmath}	
\usepackage{multirow}
\usepackage{diagbox}
\usepackage[table,xcdraw]{xcolor}

\usepackage{hyperref}
\usepackage{fontawesome} 
\usepackage{orcidlink}
\usepackage{ulem}

\usepackage{caption}
\usepackage{subfigure}

\title[GW detector anisotropy]{Detector induced anisotropies on the angular distribution of gravitational wave sources and opportunities of constraining horizon scale anisotropies}

\author[Li et al]{
Mingzheng Li$^{1,2}$\thanks{E-mail: limz01@sjtu.edu.cn} \orcidlink{0000-0002-3603-8532},
Pengjie Zhang$^{1,3,2}$\thanks{E-mail: zhangpj@sjtu.edu.cn} \orcidlink{0000-0003-2632-9915},
Wen Zhao$^{4,5}$\thanks{E-mail: wzhao7@ustc.edu.cn} \orcidlink{0000-0002-1330-2329}
\\
$^{1}$ Department of Astronomy, School of Physics and Astronomy, Shanghai Jiao Tong University, Shanghai, 200240, P.R.China\\
$^{2}$ Key Laboratory for Particle Astrophysics and Cosmology (MOE) / Shanghai Key Laboratory for Particle Physics and Cosmology, P.R.China\\
$^{3}$ Tsung-Dao Lee Institute, Shanghai Jiao Tong University, Shanghai, 200240, P.R.China\\
$^{4}$ School of Astronomy and Space Sciences, University of Science and Technology of China, Hefei, 230026, P.R.China\\
$^{5}$ Department of Astronomy, University of Science and Technology of China, Hefei, 230026, P.R.China
}

\date{Accepted XXX. Received YYY; in original form ZZZ}

\pubyear{2023}

\begin{document}
\label{firstpage}
\pagerange{\pageref{firstpage}--\pageref{lastpage}}
\maketitle

\begin{abstract}
The cosmological principle has been verified using electromagnetic observations. However its verification with high accuracy is challenging due to various foregrounds and selection effects, and possible violation of the cosmological principle has been reported in the literature. In contrast, gravitational wave (GW) observations are free of these foregrounds and related selection biases. This may enable future GW experiments to test the cosmological principle robustly with full sky distribution of millions of standard bright/dark sirens. However, the sensitivities of GW detectors are highly anisotropic, resulting in significant instrument induced anisotropies in the observed GW catalogue. We investigate these instrumental effects for 3rd generation detector networks in term of multipoles $a_{\ell m}$ of the observed GW source distribution, using Monte Carlo simulations. (1) We find that the instrument induced anisotropy primarily exists at the $m=0$ modes on large scales ($\ell \lesssim 10$), with amplitude $\langle |a_{\ell 0}|^2 \rangle \sim 10^{-3}$ for two detectors (ET-CE) and $\sim 10^{-4}$ for three detectors (ET-2CE). This anisotropy is correlated with the sky distribution of signal-to-noise ratio and localization accuracy. Such anisotropy sets a lower limit on the detectable cosmological $a_{\ell 0}$. (2) However, we find that the instrument induced anisotropy is efficiently cancelled by rotation of the Earth in $m\neq 0$ components of $a_{\ell m}$. Therefore $a_{\ell m}$ ($m\neq 0$) are clean windows to detect cosmological anisotropies. (3) We investigate the capability of 3rd generation GW experiments to measure the cosmic dipole. Through Monte Carlo simulations, we find that cosmic dipole with an amplitude of $\sim 10^{-2}$ reported in the literature can be detected/ruled out by ET-CE and ET-2CE robustly, through the measurement of $a_{11}$. 
\end{abstract}




\section{Introduction} \label{sec:intro} 

The cosmological principle is a fundamental assumption in cosmology, stating that the universe is homogeneous and isotropic. The motion of the observer in the cosmic rest-frame causes a dipole anisotropy in the cosmic microwave background (CMB) temperature sky map due to the Doppler effect, as well as the dipole in the galaxy catalogue. If the cosmological principle is valid, these two dipoles should be consistent with each other. However, several analyses of radio galaxies and quasars have reported discrepancies around $5\sigma$ or more (e.g., \citet{Bengaly2018,2021ApJ...908L..51S} and review article by \citet{2022NewAR..9501659P,2023CQGra..40i4001K}; \citet{Secrest2022,Aluri2022}). These results may suggest an intrinsically anisotropic distribution of matter in the universe, challenging the standard Lambda cold dark matter cosmology.

Electromagnetic (EM) observations suffer from various imaging systematics, such as Galactic foreground (e.g., dust extinction), star-galaxy misidentification, and atmospheric distortion (seeing and airmass). These effects induce fake fluctuations in galaxy number density and are challenging to accurately model and remove \citep{Xu2022}. These effects have a particularly strong influence on large scales, where the Galactic foreground is an issue at all EM frequencies. Thus, it is crucial to understand such large-scale observational effects in order to detect cosmic anisotropy robustly. In contrast to light, gravitational waves (GWs) are transparent to our galaxy, the Earth, and its atmosphere, making GW observations appealing alternatives to EM observations. As a result, GW sources can serve as clean tracers of matter distribution on large scales. The most likely sources of GW signals are coalescing binary compact objects, including binary black holes (BBH), Binary Neutron Stars, and Neutron Star-Black Hole binaries. Since compact binaries are located in galaxies, the galaxy distribution and the large scale structure of the universe can be mapped with enough GW events and precise 3D localization.

\begin{figure}
	\includegraphics[width=\columnwidth]{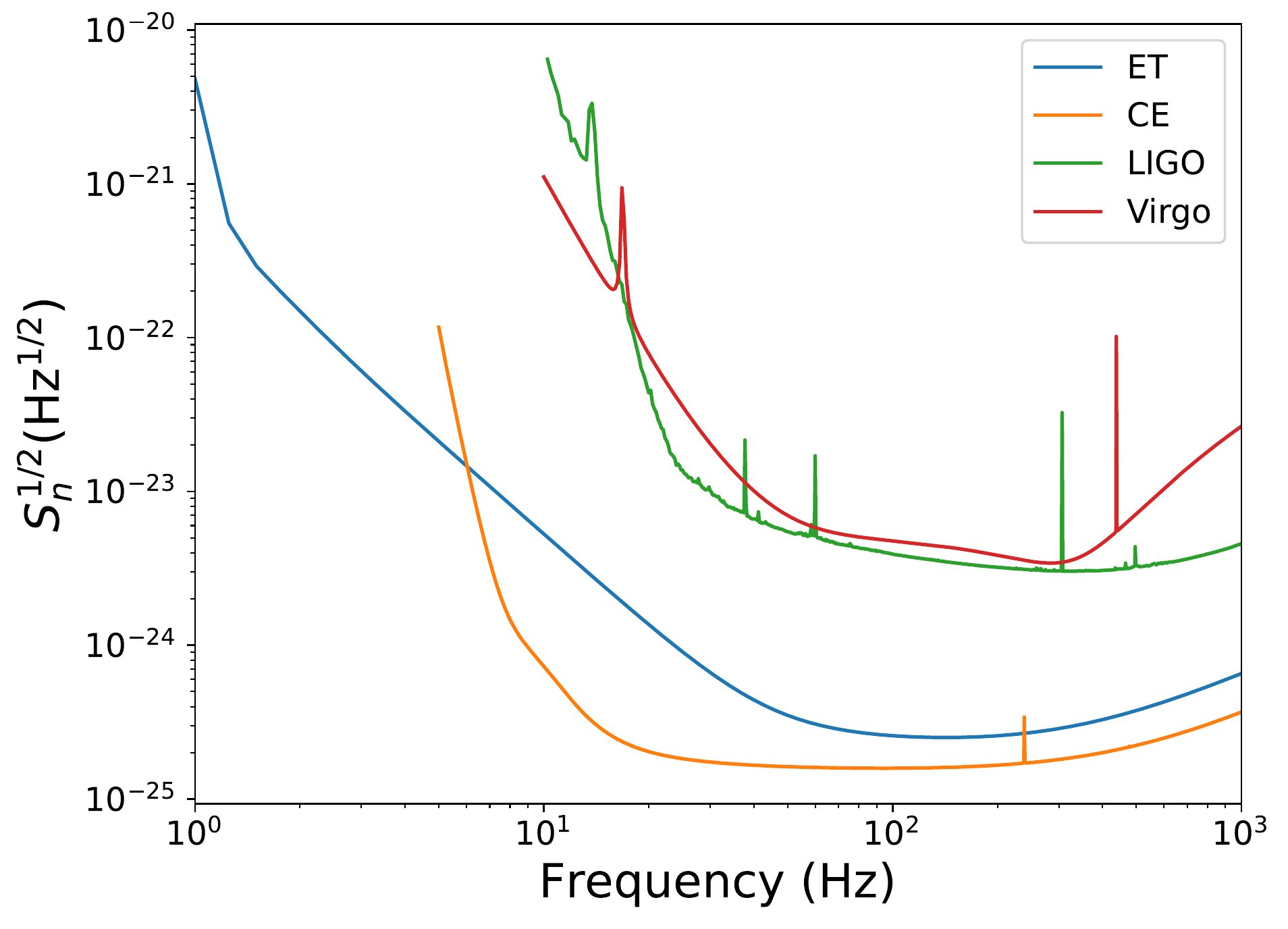}
    \caption{Noise ASD of 2nd generation (LIGO and Virgo) and 3rd generation (ET and CE) detectors. We use data of noise curves from \citet{Abbott2020a} for 2nd generation detectors and from \citet{Abbott2017} for 3rd generation detectors.}
    \label{fig:psd}
\end{figure}

The LIGO-Virgo-KAGRA collaboration has confirmed over 100 GW events after the latest O3b release \citep{TheLIGOScientificCollaboration2021}. The number of detections is still increasing as observations continue. The currently running GW experiments, such as Advanced LIGO \citep{Aasi2015b}, Advanced Virgo \citep{Accadia2012b}, and KAGRA \citep{KAGRACollaboration2019b}, are using 2nd generation detectors. While proposed 3rd generation GW detectors, such as the Cosmic Explorer (CE) \citep{Reitze2019} and the Einstein Telescope (ET) \citep{Maggiore2019}, are designed with longer arm lengths and improved control on the quantum noise, resulting in significant improvements on sensitivity to GW signals. Figure \ref{fig:psd} shows the amplitude spectral density (ASD) of the noise of CE, ET, LIGO, and Virgo. The noise amplitudes of 3rd generation detectors are at least an order-of-magnitude lower than that of 2nd generation detectors, with more extended frequency coverage. Orders of magnitude more GW events with larger distances and smaller masses will be detected with 3rd generation detectors \citep{Zhao2018}.

Several studies have investigated the constraining power of cosmic anisotropy through GW observation. \citet{Galloni2022} explored the possibility of using the Cosmological GW Background, detected by the proposed spaceborne detectors, including Laser Interferometer Space Antenna (LISA) and Big Bang Observer (BBO), to investigate the CMB hemisphere power asymmetry anomaly. \citet{Kashyap2022} studied the anisotropy of BBH mass distribution caused by observer motion using currently detected events from LIGO-Virgo. The result is consistent with isotropy. \citet{Essick2022} performed hierarchical Bayesian inference on $63$ detected events from the O3 observation run of the LIGO-Virgo-KAGRA collaboration, also found no evidence of anisotropy. \citet{Mastrogiovanni2022} investigated the dipole anisotropy induced in GW detection by the motion of the observer, taking the assumption that the response and localization accuracy are constant. 

An issue awaiting thorough investigation is the anisotropy introduced by GW detectors. The detector response is strongly direction-dependent, resulting in strongly anisotropic sensitivity on different regions of the sky and inducing anisotropies in the observed GW source distribution. Therefore, it is essential to quantify and mitigate the impact of GW detector anisotropy on the detection of cosmic anisotropy. We employ MC simulations to generate isotropically distributed BBH sources, then apply instrumental effects to each source and analyse the anisotropy in the resulting detection catalogue. We also identify clean windows $a_{\ell m}$ ($m\neq 0$) for cosmic anisotropy detection, in the spherical harmonic space and with the aid of Earth rotation. The detector induced anisotropy depends on positions and orientations of the GW detector/network. Generally, this anisotropy differs in directions form the cosmological anisotropies we are interested in, such as the cosmic dipole. Therefore we can use the uncontaminated $a_{\ell m}$ ($m\neq 0$) modes for our search. We take the cosmic dipole as an example. Since researches of radio galaxies suggest that a dipole different from the CMB dipole exists for large scale galaxy distribution, we simulate anisotropically distributed sources whose sky distribution follows this dipole. We discover that GW events can be used to detect cosmic dipoles through clean windows unaffected by instrument induced anisotropy.

This article is structured as follows. In section \ref{sec:WaveformAndFisher}, we provide a brief introduction on the method we use for GW data analysis and our techniques for numerical evaluating the Fisher matrix. In section \ref{sec:InstrumentAnisotropy}, we describe the configuration of our Monte Carlo (MC) simulations and present our results on the study of instrumental anisotropy. In section \ref{sec:DipoleDetectability}, we simulate GW sources with intrinsic dipole distributions and demonstrate the ability to detect and constrain these dipoles using the GW catalogue. Finally, in section \ref{sec:conclusion}, we summarize our results and present our conclusions.

\section{Gravitational Wave Data Analysis and Fisher Matrix} \label{sec:WaveformAndFisher}

\subsection{Detector Response to gravitational wave Signal} \label{sec: GWResponse}

GWs have two independent polarizations, $h_{+}$ and $h_{\times}$. Each polarization push and squeeze space in different way, and generate different strain signal on interferometers. The total strain signal is given by
\begin{equation} \label{eqn:strainResponse}
    h = F_+h_+ + F_\times h_\times\ .
\end{equation}
Here $F_+$ and $F_\times$ are functions of relative directions of GW source and detector arms \citep{Nishizawa2009},
\begin{equation}
\begin{aligned} \label{eqn:responseInAngles}
F_{+}(\theta, \phi, \psi)=& \frac{1}{2}\left(1+\cos ^{2} \theta\right) \cos 2 \phi \cos 2 \psi  -\cos \theta \sin 2 \phi \sin 2 \psi \\
F_{\times}(\theta, \phi, \psi)=&-\frac{1}{2}\left(1+\cos ^{2} \theta\right) \cos 2 \phi \sin 2 \psi -\cos \theta \sin 2 \phi \cos 2 \psi
\end{aligned}
\end{equation}
where $\theta,\phi$ are direction angles in the detector frame and $\psi$ is the polarization angle. Clearly the signal detected by the interferometers varies with the direction of GW sources. 

The detected strain data $h(t)$, as function of time, is often processed in frequency domain $h(f)$, through the Fourier transformation. The signal-to-noise ratio (SNR) of GW signal $h(f)$ is 
\begin{equation} \label{eqn:SNRDef}
    \rho = (h,h)\ .
\end{equation}
Here the inner product of two frequency domain signal $a(f)$ and $b(f)$ is defined as
\begin{equation} \label{eqn:innerProduct}
(a,b) = 4\mathrm{Re} \int_0^\infty \frac{a(f) b^*(f)}{S_n(f)} \mathrm{d}f\ .
\end{equation}
$S_n(f)$ is the noise power spectrum of the detector. For a detector network containing multiple detectors, the network SNR is defined as \citep{Zhao2018}
\begin{equation} \label{eqn:SNRNetDef}
    \rho = \sqrt{\sum_i \rho_i^2}
\end{equation}
where $\rho_i$ is the SNR of each detector in the network.

\subsection{Fisher Matrix Method in gravitational wave Data Analysis} \label{sec:FisherMatrix} 

The Fisher matrix is a commonly used tool for estimating the ability of detectors to constrain waveform parameters in GW data analysis. The elements of the Fisher matrix are defined as
\begin{equation} \label{eqn:FisherDef}
    \Gamma_{ij} = \left(\frac{\partial h}{\partial \lambda_i}, \frac{\partial h}{\partial \lambda_j} \right)
\end{equation}
where $\{\lambda\}$ represents the parameters of the GW signal, such as binary masses, distance, and the direction on the sky. In the high-SNR approximation, the posterior distribution of the parameters follows a multidimensional Gaussian distribution, with the covariance matrix being the inverse of the Fisher matrix. The variance of parameter $\lambda_i$ can then be calculated as
\begin{equation} \label{eqn:paraVariance}
    \langle \Delta \lambda_i^2 \rangle = \Gamma^{-1}_{ii}\ .
\end{equation}
We use the definition of localization accuracy in \citet{Wen2010}
\begin{equation} \label{eqn:localizationDef}
    \Delta \Omega_s = 2 \pi|\sin \theta| \sqrt{\left\langle\Delta \theta^{2}\right\rangle\left\langle\Delta \phi^{2}\right\rangle-\langle\Delta \theta \Delta \phi\rangle^{2}}\ .
\end{equation}
$\Delta \Omega_s$ quantifies the uncertainty of source localization. The probability of a source being localized outside an error ellipse of area $\Delta \Omega$ is given by $e^{-\Delta \Omega/\Delta \Omega_s}$. 

For compact binary merger, there are totally $15$ parameters, including binary masses $m_1$ and $m_2$,the luminosity distance $D_L$, the orbital inclination $\iota$, the phase $\varphi$, the arrival time $t$, the direction angles $\theta,\phi$, the polarization angle $\psi$, and six components of the spin vectors of the binary. Calculating the Fisher matrix for all $15$ parameters for millions of sources is highly computationally expensive. Since the spin has little impact on source localization \citep{Singer2016,Hu2021}, we choose to neglect the spin in this study in order to speed up the calculation. Therefore $\lambda=(m_1,m_2,D_L, \iota, \varphi, t, \theta, \phi, \psi)$. 

We calculate the Fisher matrix using numerical differentiation under the following approximation
\begin{equation} \label{eqn:numDiff}
    \frac{\partial h}{\partial \lambda} \approx \frac{h(\lambda+s)-h(\lambda-s)}{2s}
\end{equation}
where $s$ is a small step, here we set $s=10^{-7}$. The values of parameters used in the calculation are expressed in following units: solar mass for binary mass, $\mathrm{Mpc}$ for distance, seconds for time, and radians for angles. Since we need to calculate Fisher matrix for BBHs from all direction on the sky, we develope a method to handle the singularity of spherical coordinate at the north and south pole. In detail, for a source located at direction $(\theta,\phi)$, we perform a rotation $R$ to align it with the $x$-axis, hence $(\theta',\phi')=(\frac{\pi}{2},0)$. We then calculate the numerical differentiation in the rotated $(\theta',\phi')$ frame, eventually the differentiation on direction angles becomes 
\begin{equation} \label{eqn:numDiffOnDirection}
    \frac{\partial h}{\partial \lambda'} \approx \frac{h(R^{-1}(\lambda'+s))-h(R^{-1}(\lambda'-s))}{2s}\ .
\end{equation}
The rotation matrix $R$ is stored along with the Fisher matrix. We calculate the posterior of direction in the rotated frame first, then perform rotation $R^{-1}$ to obtain the correct posterior in the origin frame.

\section{The Detector Induced Anisotropy} \label{sec:InstrumentAnisotropy}
\begin{figure*}
    \centering
    \subfigure[$m=0$ modes \label{fig:Uni2DetSubM0}]{\includegraphics[width=0.45\textwidth]{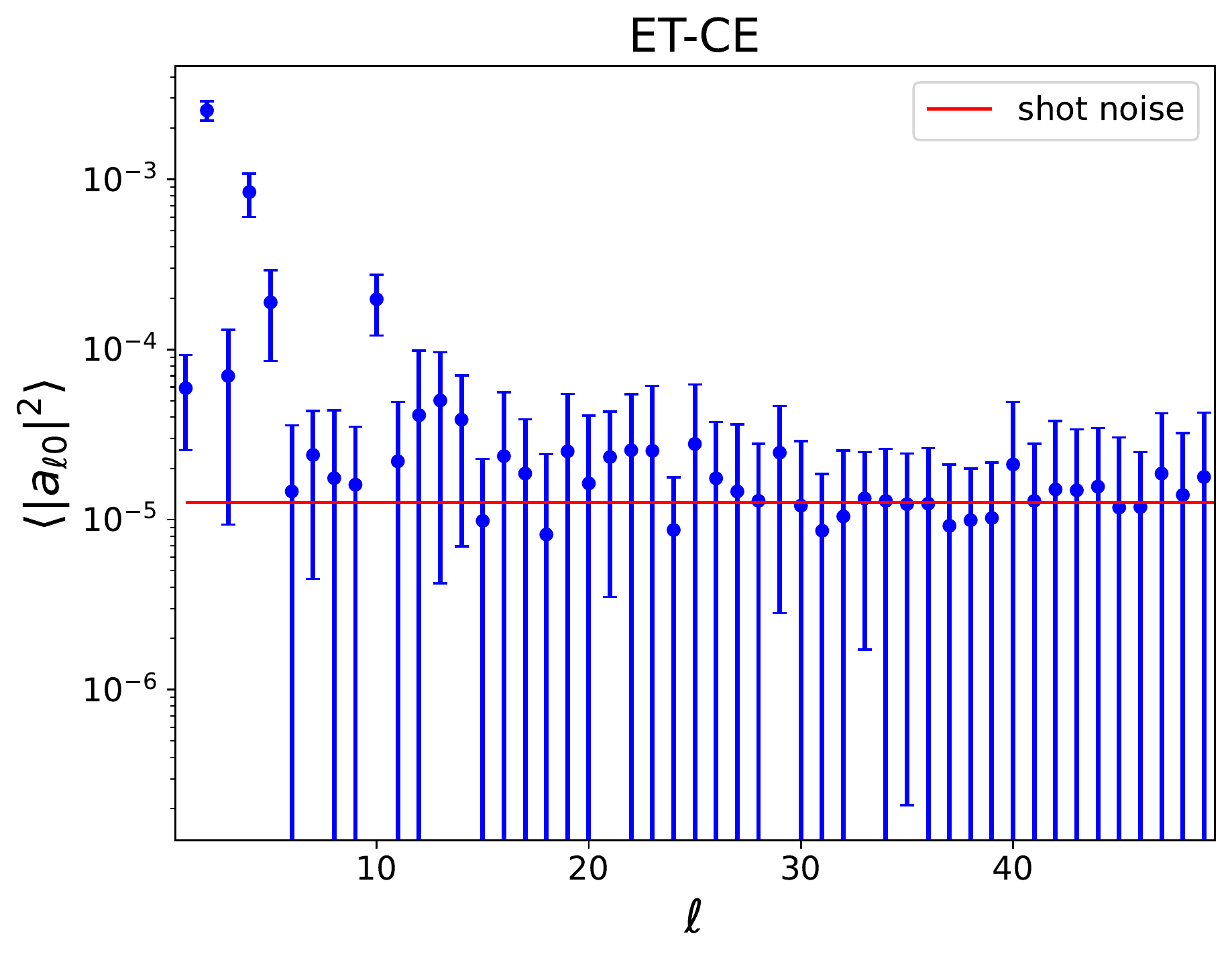}}
    \hspace{0.05\textwidth}
    \subfigure[$m \neq 0$ modes \label{fig:Uni2DetSubMNz}]{\includegraphics[width=0.45\textwidth]{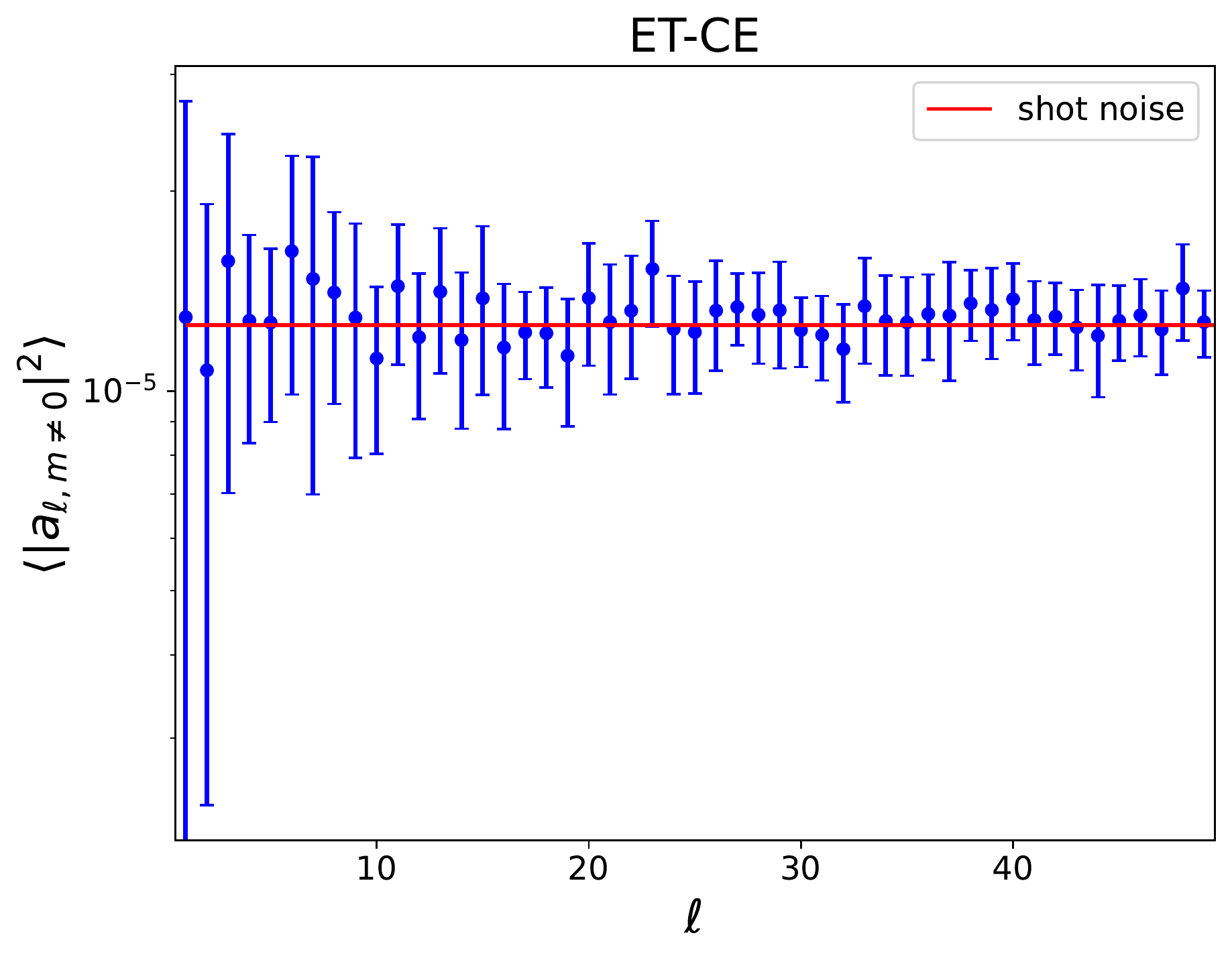}}
    \caption{The angular power spectra of the ET-CE detector induced anisotropy. The $1\sigma$ error bars are estimated using $16$ realizations. Subfigure \ref{fig:Uni2DetSubM0} shows the results of $C_{\ell 0}$ up to order $\ell=50$, while subfigure \ref{fig:Uni2DetSubMNz} shows the results of $C_{\ell, m\neq 0}$. The red line shows power spectrum of shot noise $C_{\ell 0}^{\rm shot}$. The detector induced anisotropy is statistically significant above the shot noise at $\ell<10$. But the $m\neq 0$ modes are free of this effect, due to the Earth rotation.}
    \label{fig:Uni2Det}
\end{figure*}

\subsection{Simulation Configuration} \label{sec:MCConfig}

Even if the GW sources are isotropically distributed on the sky, the anisotropic detection ability of the detectors can lead to an anisotropic distribution in the GW catalogue. Two factors contribute to this anisotropy. The first one is the anisotropic response of the detector, lower sensitivity in certain directions results in fewer number of detections. The second is the anisotropic localization accuracy, which causes a larger scatter from the true source location in directions with worse localization accuracy, also leading to fewer detections.

To study this anisotropy, we employ MC method to generate $10^6$ isotropically distributed BBH samples, which could be achieved in a few years for 3rd generation detectors \citep{Zhao2018,Yang2022}. For each BBH, we randomly sample the waveform parameters and calculate the SNR and the Fisher matrix. The observed parameters are sampled from a multidimensional Gaussian distribution with mean values equal to the true value and covariance equal to the inverse of the Fisher matrix. The SNR threshold is set to $10$, and events with lower SNR are considered undetected.

We choose the BBH population based on current GW observations. \citet{TheLIGOScientificCollaboration2020} provides models for the binary mass distribution of BBH mergers based on Bayesian analysis of the gravitational-wave transient catalogue (GWTC-2). We adopt the Power Law + Peak Model with the most probable model parameters, since this model has the highest Bayesian factor and is therefore the most favored. The distance prior is set to ensure that the sources are uniformly distributed in comoving space, with a redshift range of $0.1 \le z \le 2$. The direction of the orbit plane is set to be uniformly distributed, resulting in a $\sin$ prior of the orbit inclination angle.

We develop a {\tt PYTHON} package {\tt GWSKY} that performs the entire workflow from MC simulation to generation and analysis of the observed catalogue. Several modules of {\tt GWSKY} are based on the package {\tt BILBY} \citep{Ashton2019}. Source code of {\tt GWSKY} is available on GitHub \faGithub~ \url{https://github.com/mzLi01/gwsky}.

\subsection{Quantifying the Instrumental Anisotropy} \label{sec:AniModesDef}

With the observed catalogue, one can calculate the overdensity of GW sources as a function of the direction $\hat{n}$
\begin{equation} \label{eqn:overdensityDef}
\begin{aligned}
\delta (\hat{n}) &= \frac{n_{\rm GW}}{\Bar{n}_{\rm GW}} - 1  = \frac{4\pi}{N} \sum_i \delta_{\rm 2D}(\hat{n}-\hat{n}_i) - 1
\end{aligned}
\end{equation}
where $\delta_{\rm 2D}$ is the 2D Dirac delta defined on the sphere, and $\hat{n}_i$ is the observed direction of the $i$-th GW source. Here we use the equatorial coordinate, and the equatorial plane has $\theta=\pi/2$. Since the sources are isotropically distributed, the observed anisotropy is purely instrumental effect.

To quantitatively study the instrumental anisotropy, we apply spherical harmonic decomposition on the overdensity. Each $\ell\ge 1$ spherical harmonic mode corresponds to an independent anisotropy mode, larger $\ell$ modes corresponding to anisotropy on smaller scales. The coefficients $a_{\ell m}$ can be expressed as
\begin{equation} \label{eqn:almValues}
a_{\ell m} = \int \mathrm{d}^2\hat{n} \delta(\hat{n}) Y_{\ell m}^*(\hat{n}) = \frac{4\pi}{N}\sum_i Y_{\ell m}^*(\hat{n}_i) \ .
\end{equation}

We run the simulation described in Section \ref{sec:MCConfig} $16$ times to obtain $16$ realizations of the catalogue. For each realization, we calculate the spherical harmonic coefficients up to $\ell=50$ to study the components of the instrumental anisotropy. We mainly focus on the $m=0$ modes, since $m\neq 0$ modes correspond to anisotropies in the longitude direction, which are averaged out due to rotation of the Earth. Specifically, the instrumental effects only depend on the relative direction between the source and the detector, and the random arrival time of the sources results in a shift in the longitude direction of the relative source direction. Therefore, for a given $\theta$, the instrumental anisotropy $\delta(\theta,\phi)$ is approximately the average value of every $\phi$ in $[0,2\pi]$, which means $\delta(\theta,\phi) \approx \delta(\theta)$ and hence $a_{\ell m}\approx 0$ for $m\neq 0$ modes.

The angular power spectrum
\begin{equation} \label{eqn:APSDef}
C_{\ell 0}\equiv \langle|a_{\ell 0}|^2\rangle\ .
\end{equation}
is calculated by taking average over all $16$ realizations. The standard errors among the realizations are also calculated. For a comparison, the shot noise power spectrum caused by the discrete GW source distribution is 
\begin{equation}
    C_{\ell 0}^{\rm shot}=\frac{4\pi}{N_{\rm GW}}\ .
\end{equation}
Here $N_{\rm GW}$ is the total number of GW sources for the anisotropy measurement. For comparison, the angular power spectrum of $m\neq 0$ modes $C_{\ell, m\neq 0}$ is also calculated, 
\begin{equation} \label{eqn:APSMnzDef}
    C_{\ell, m\neq 0}\equiv \frac{\langle\sum_{m=1}^{\ell} |a_{\ell m}|^2\rangle}{\ell}\ .
\end{equation}

\subsection{Results for detector network ET-CE and ET-2CE} \label{sec:UniNetResults}

The result of \citet{Zhao2018} has already demonstrated that for single detector, the localization error for GW sources can reach the magnitude comparable to area of the entire sky ($\gtrsim 10^4 \deg^2$), making GW observations no longer able to provide information of cosmic anisotropy. Therefore, in this study, we only consider detector networks consisting of multiple detectors.
\begin{figure*}
    \centering
    \subfigure[$m=0$ modes]{\includegraphics[width=0.45\textwidth]{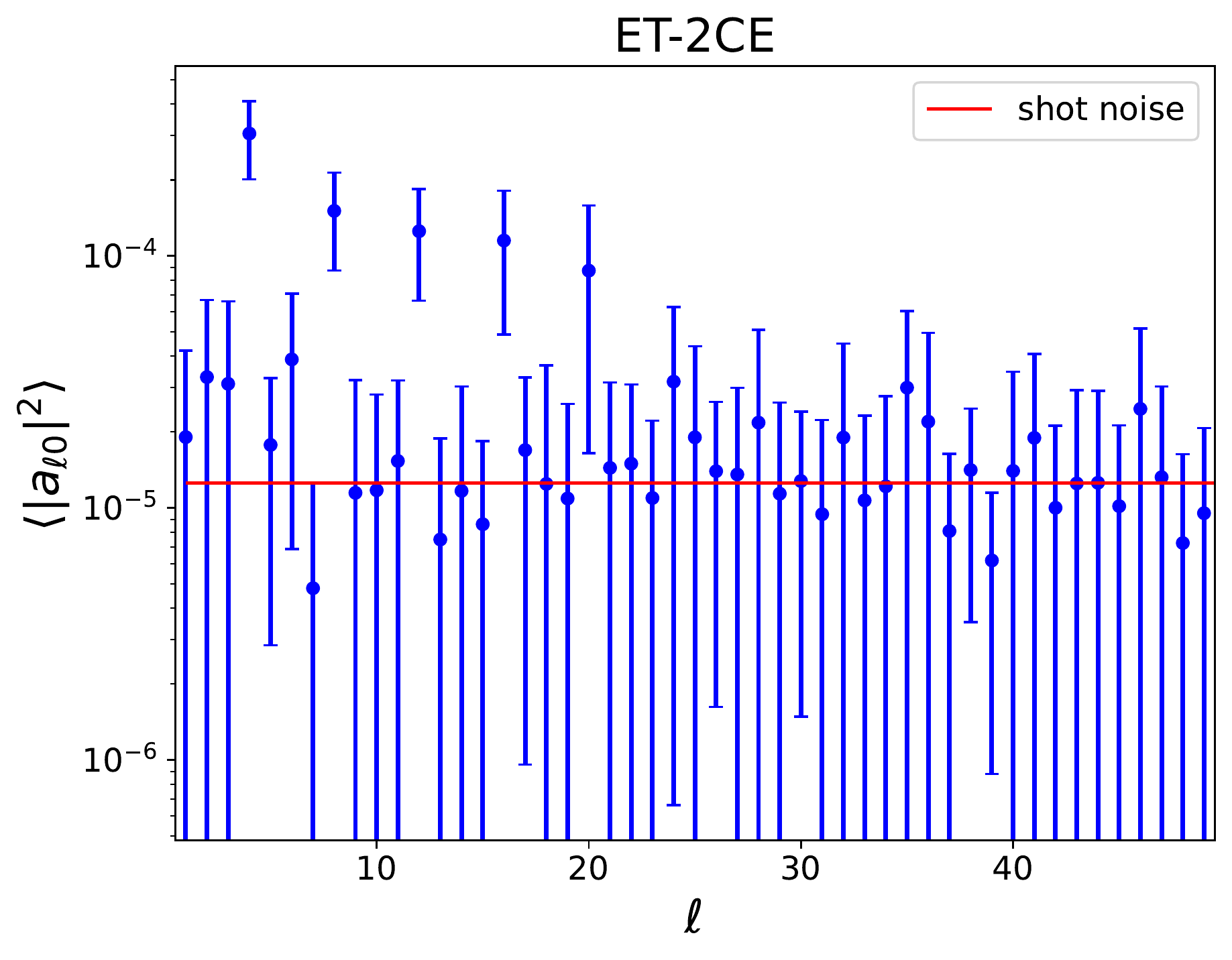}}
    \hspace{0.05\textwidth}
    \subfigure[$m \neq 0$ modes]{\includegraphics[width=0.45\textwidth]{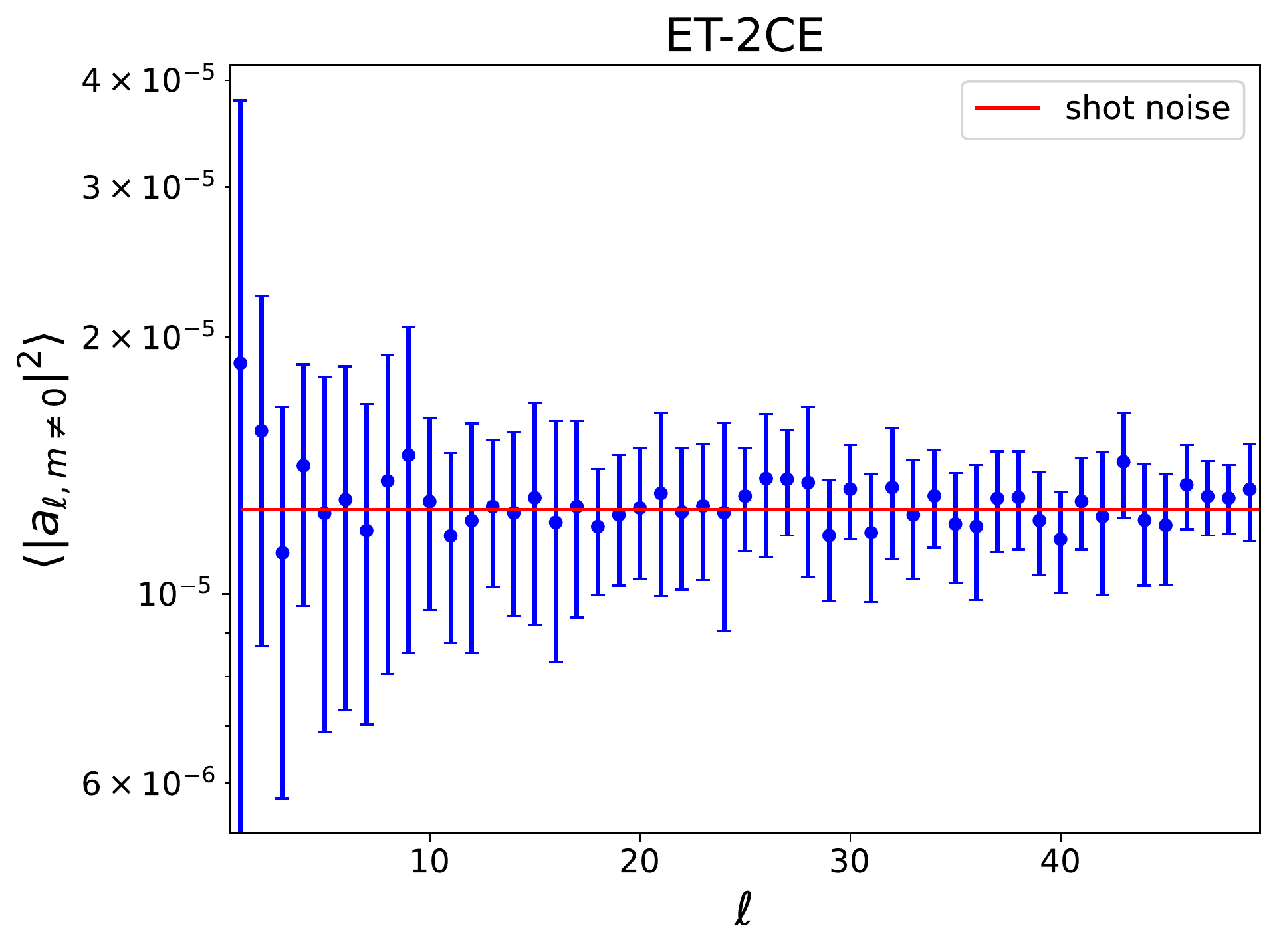}}
    \caption{Same as Figure \ref{fig:Uni2Det}, but for the ET-2CE network. The overall anisotropy of the $m=0$ mode is a factor of $10$ weaker than the ET-CE case. The $m\neq 0$ mode is again consistent with Poisson fluctuations. Therefore these $m\neq 0$ modes are clean windows for detecting the cosmic anisotropies. }
    \label{fig:Uni3Det}
\end{figure*}

We investigate two network configurations of ET and CE, adopting the location of the detectors in \citet{Zhao2018}.
\begin{itemize}
    \item The ET-CE configuration. It includes one ET located at the Virgo site and one CE located at the LIGO site.
    \item The ET-2CE configuration. It includes one additional CE located in Australia. 
\end{itemize}

Figure \ref{fig:Uni2Det} and Figure \ref{fig:Uni3Det} show the angular power spectra for anisotropy induced by the ET-CE and ET-2CE network. The detector induced anisotropy beyond the Poisson fluctuations is statistically significant for $a_{\ell 0}$ at $\ell<10$ for both network configurations. In contrast, the $a_{\ell,m\neq 0}$ modes are consistent with the Poisson fluctuations. Another finding is that detector induced anisotropies in the $m=0$ mode decrease by a factor of $10$ from ET-CE to ET-2CE. Therefore the major conclusion here is that all the $m=0$ modes are heavily contaminated by the detector induced anisotropy, while the $m\neq 0$ modes are clean windows for detection of cosmic anisotropy. 

We believe that the instrumental anisotropy is caused by the anisotropic response and localization accuracy, as discussed in Section \ref{sec:MCConfig}. To gain further insight into its origin, we calculate its cross-correlation spectrum with the SNR sky map and the localization accuracy sky map. The sky maps are calculated by putting a BBH source at different directions on the sky and calculating SNR and localization accuracy for each direction. The other parameters of the source are fixed since we are mainly interested in the properties of the detector network in this calculation. We adopt the BBH primary mass $m_1=10 M_\odot$, mass ratio $r=0.9$ and redshift $z=1$. This is a simplification, but sufficient for our purpose of demonstrating the connection between anisotropy in the detected GW distribution and SNR/localization accuracy. We divide the sky into $49152$ equal area pixels using {\tt HEALPIX} with $N_\mathrm{side}=64$, and put one source in the center of each pixel to obtain the sky maps. 

We quantify the possible connections with the cross-correlation coefficient between two maps $f$ and $g$. 
\begin{equation}
    \label{eqn:Normed2DCor}
    r(\ell_{\rm max})\equiv \frac{\sum_{\ell=0}^{\ell_\mathrm{max}} f_{\ell m} g^*_{\ell m}}{\sqrt{\left(\sum_{\ell=0}^{\ell_\mathrm{max}} f_{\ell m} f^*_{\ell m}\right) \left(\sum_{\ell=0}^{\ell_\mathrm{max}} g_{\ell m} g^*_{\ell m}\right)}}\ .
\end{equation}
$f_{\ell m}$ and $g_{\ell m}$ are the spherical coefficients of $f$ and $g$. In the limit of $\ell_{\rm max}\rightarrow \infty$, 
\begin{equation}
    r(\ell_{\rm max}\rightarrow \infty)\rightarrow \frac{\langle fg\rangle}{\sqrt{\langle f^2\rangle\langle g^2\rangle}}\ .
\end{equation}

\begin{figure}
	\includegraphics[width=\columnwidth]{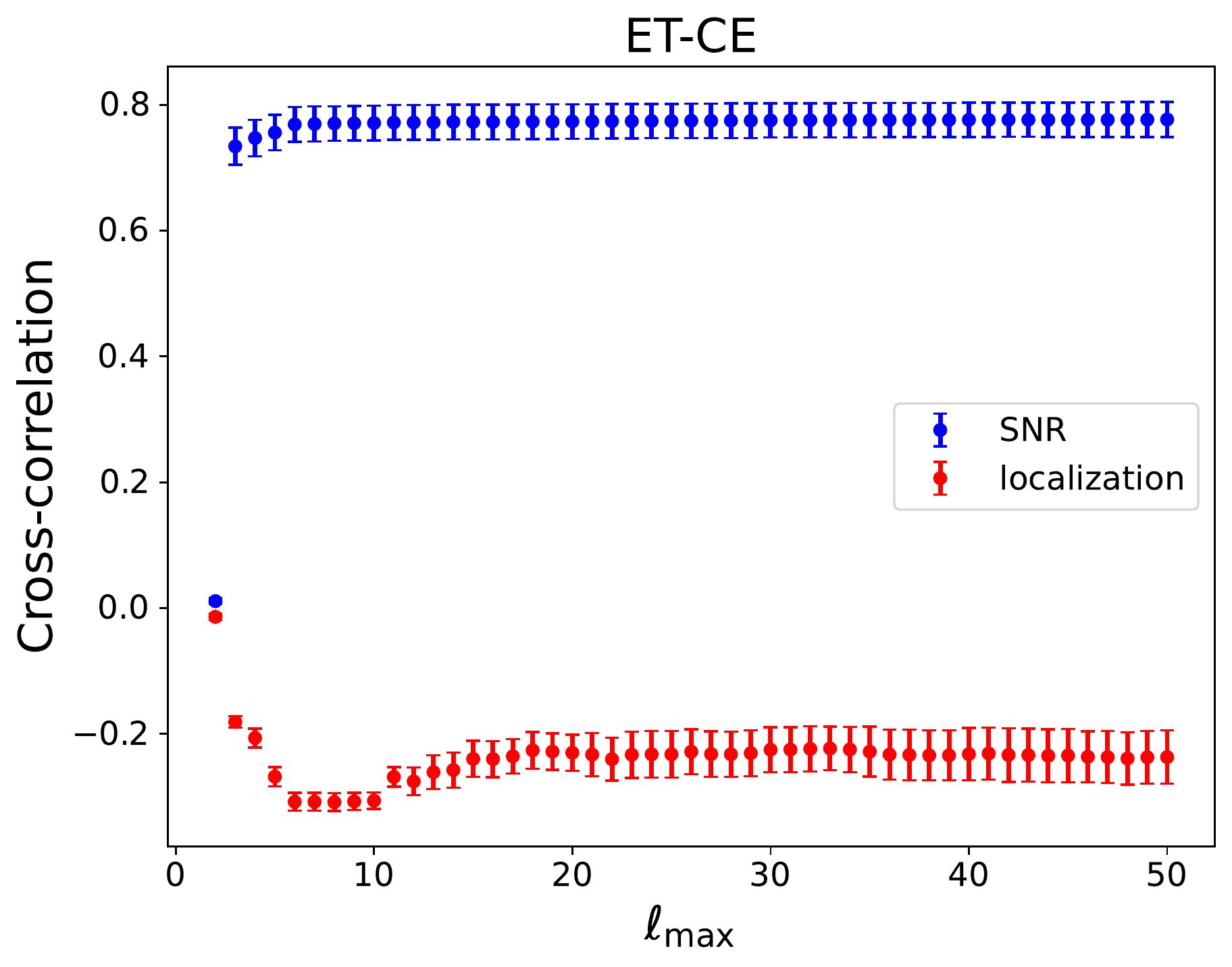}
    \caption{Cross-correlations $r(\ell_{\rm max})$ of instrumental anisotropy with SNR and with localization accuracy sky map for ET-CE network. The $1\sigma$ error bars are estimated using $16$ realizations.}
    \label{fig:SHCor2Det}
\end{figure}

\begin{figure}
	\includegraphics[width=\columnwidth]{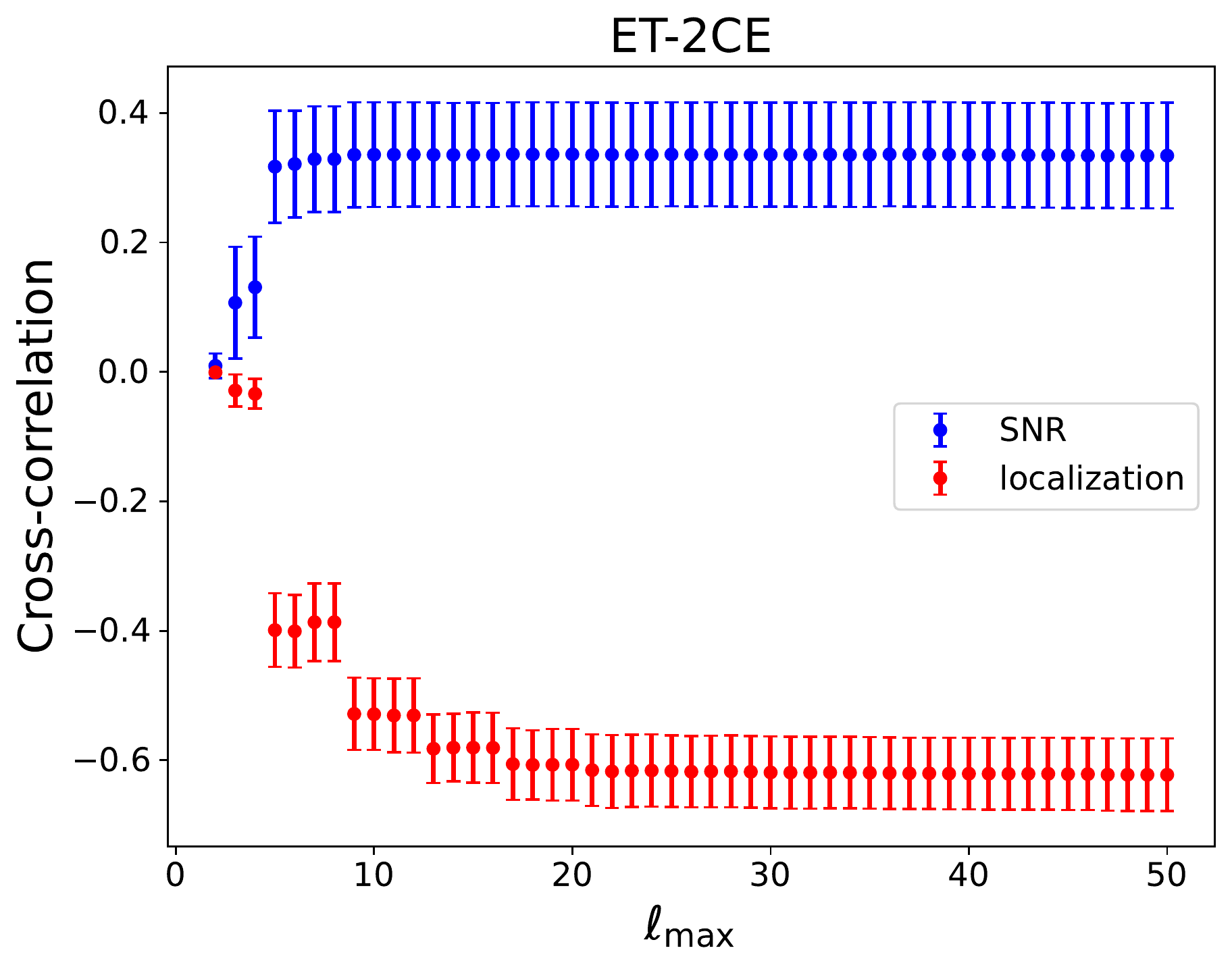}
    \caption{Same as Figure \ref{fig:SHCor2Det}, but for the ET-2CE network.}
    \label{fig:SHCor3Det}
\end{figure}

Figure \ref{fig:SHCor2Det} and Figure \ref{fig:SHCor3Det} show $r(\ell_{\rm max})$ with $\ell_{\rm max}$ ranging from $2$ to $50$, for ET-CE and ET-2CE respectively. From Figure \ref{fig:SHCor2Det} we notice that the ET-CE instrumental anisotropy strongly correlates with the SNR anisotropy, while for ET-2CE, the correlation with localization anisotropy is stronger than SNR anisotropy, as displayed in Figure \ref{fig:SHCor3Det}. This can be explained by considering that the ET-CE network has relatively worse ability to detect BBHs. Namely a BBH of the adopted property for the S/N estimation has a significant chance to be undetected by ET-CE. The resulting SNR anisotropy dominates the instrumental anisotropy. Adding another detector to the ET-CE network improves the localization of BBHs, resulting in a decrease of anisotropy on small scales (large $\ell$), while low $\ell$ modes remain due to the existence of global symmetries. The distribution of spherical harmonic components of localization anisotropy is therefore more similar to the instrumental anisotropy. On the other hand, the increase in the total response of the network reduces the fraction of BBHs that are not detected ($\mathrm{SNR}<10$), making the SNR cut less significant, so the correlation with SNR drops. As a result, adding a detector to the ET-CE network enhances the contribution of the localization accuracy to the instrumental anisotropy.

\section{Detectability of Cosmic Dipole} \label{sec:DipoleDetectability}
The existence of clean windows free of detector induced anisotropy allows us to probe cosmic anisotropies. A prominent example is the cosmic dipole. The CMB dipole has been measured independently through the leading order effect ($\ell=1$ mode in the CMB temperature map), and through higher order effect (the induced correlation between $\ell$ modes and $\ell\pm 1$ modes, \citet{2014A&A...571A..27P}). The agreement of the two measurements confirms that the Doppler motion of the Solar system alone is the main cause of the CMB dipole, while the the intrinsic cosmic dipole can be considered negligible. This conclusion supports the cosmological principle \citep{2014A&A...571A..27P}. However, several studies based on galaxy and AGN distributions have reported a dipole in the galaxy/AGN distribution with amplitude much larger than the CMB dipole, but roughly at the same direction (Table \ref{tab:DipoleRef}). 

In this section, we investigate the detectability of the possible dipole reported. We assume the directions of BBHs follow the cosmic dipole, while the instrumental anisotropies are calculated in the same way as in Section \ref{sec:InstrumentAnisotropy}. The cosmic dipole can be detected if we can observe significant $a_{1m}\neq 0$ signal. Specifically, we sample the directions of BBHs using a dipole distribution on the sky and analyse the statistical significance of $a_{11}$, which is free of detector induced anisotropies. For the dipole parameters, we adopt the results of \citet{Secrest2022}, which is the latest study, and the first joint analysis of radio galaxies and quasars. We set the dipole direction as $(l,b)=(233.0^\circ, 34.4^\circ)$, and test three dipole amplitudes $\mathcal{D}=0.005,0.01,0.02$. $\mathcal{D}$ is defined such that the dipole contribution of the overdensity is $\delta(\hat{n})=\mathcal{D} \cos\Delta\theta$, where $\Delta\theta$ is the angle between $\hat{n}$ and dipole direction. The distribution of the other parameters are the same as in Section \ref{sec:MCConfig}. For each dipole amplitude, we calculate $16$ realizations, to compare with the $16$ realizations calculated in Section \ref{sec:InstrumentAnisotropy}.

\begin{figure}
	\includegraphics[width=\columnwidth]{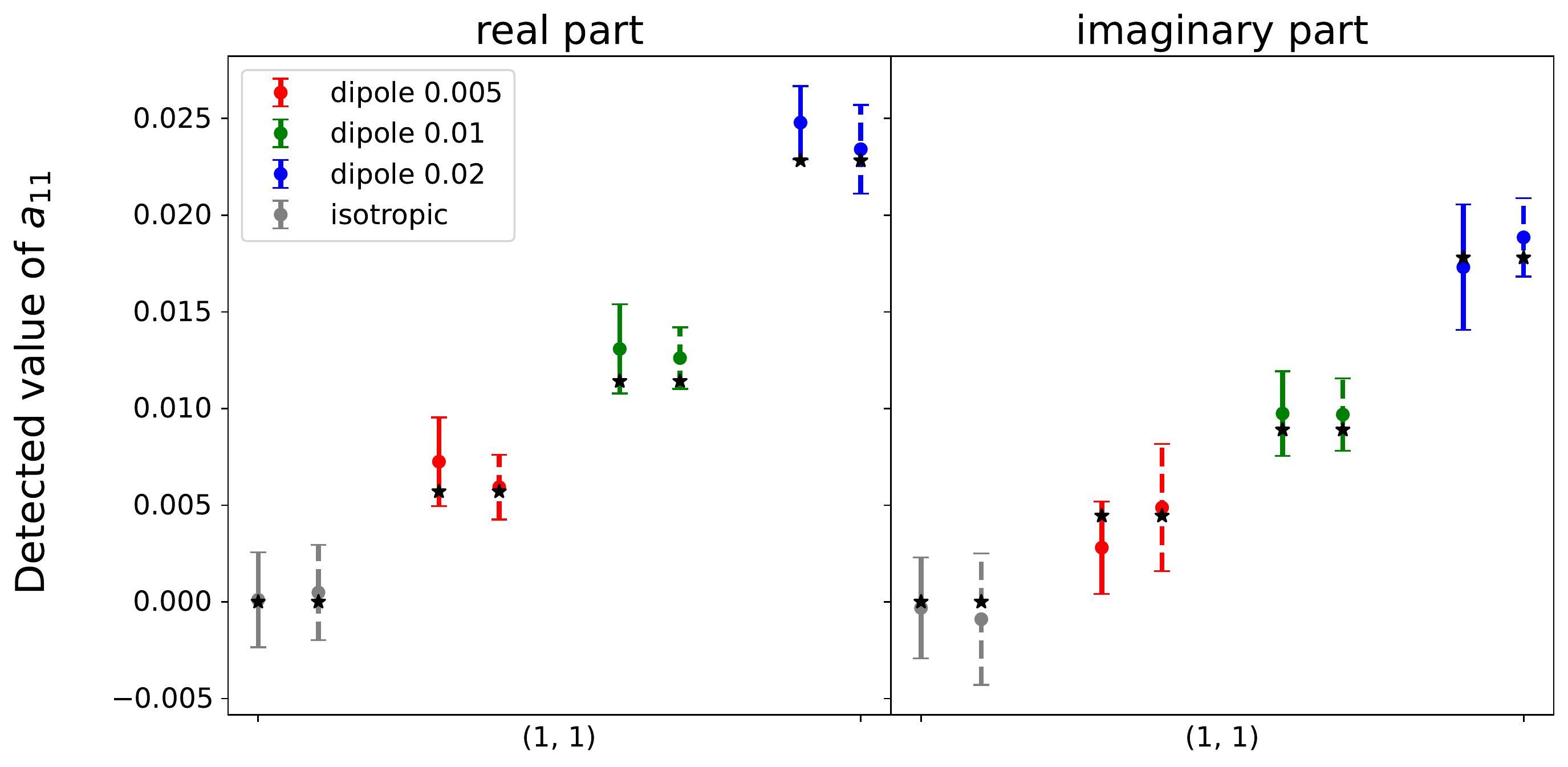}
    \caption{Measured $a_{11}$ for simulated  GW sources with input dipole amplitude $\mathcal{D}=0,0.005,0.01,0.02$. The $1\sigma$ error bars are estimated using 16 realizations. The dashed error bars represent ET-2CE results, while solid error bars represent ET-CE results. The input values of $a_{11}$ are marked with the black star symbol on each error bar. }
    \label{fig:Dipole}
\end{figure}

\begin{table*}
\begin{tabular}{cccc}
\hline
Amplitude         & Direction $(l,b)$                                                                & Survey                & Reference                            \\ \hline
$0.021 \pm 0.006$ & $\left(244.69^{\circ} \pm 27.00^{\circ}, 41.18^{\circ} \pm 29.00^{\circ}\right)$ & \multirow{7}{*}{NVSS} & \citet{Blake2002}                    \\
$0.021 \pm 0.005$ & $\left(252.22^{\circ} \pm 10.00^{\circ}, 42.74^{\circ} \pm 9.00^{\circ}\right)$  &                       & \citet{Singal2011}                   \\
$0.027 \pm 0.005$ & $\left(213.99^{\circ} \pm 20.00^{\circ}, 15.30^{\circ} \pm 14.00^{\circ}\right)$ &                       & \citet{Gibelyou2012}                 \\
$0.019 \pm 0.005$ & $\left(248.47^{\circ} \pm 19.00^{\circ}, 45.56^{\circ} \pm 9.00^{\circ}\right)$  &                       & \citet{Rubart2013}                   \\
$0.010 \pm 0.005$ & $\left(256.49^{\circ} \pm 9.00^{\circ}, 36.25^{\circ} \pm 11.00^{\circ}\right)$  &                       & \citet{Tiwari2013}                   \\
$0.012 \pm 0.005$ & $\left(253.00^{\circ}, 32.00^{\circ}\right)$                                     &                       & \citet{Tiwari2015}                   \\
$0.019 \pm 0.002$ & $\left(253.00^{\circ} \pm 2.00^{\circ}, 28.71^{\circ} \pm 12.00^{\circ}\right)$  &                       & \citet{Colin2017}                    \\ \hline
$0.070 \pm 0.004$ & $\left(243.00^{\circ} \pm 12.00^{\circ}, 45.00^{\circ} \pm 3.00^{\circ}\right)$  & TGSS                  & \multirow{2}{*}{\citet{Bengaly2018}} \\
$0.023 \pm 0.004$ & $\left(253.12^{\circ} \pm 11.00^{\circ}, 27.28^{\circ} \pm 3.00^{\circ}\right)$  & NVSS                  &                                      \\ \hline\hline
$0.014 \pm 0.001$ & $\left(233.0^{\circ}, 34.4^{\circ}\right)$                                       & NVSS+WISE             & \citet{Secrest2022}                  \\ \hline
\end{tabular}
    \caption{Previous measurements of cosmic dipole obtained by number count of radio galaxies and AGNs. The directions are expressed in the Galactic coordinate. As a reference, the CMB dipole is at direction $(l,b)=(264^{\circ},48^{\circ})$ with an amplitude of $1.3\times 10^{-3}$ \citep{2014A&A...571A..27P}. Data of this table is based on Table 2 of \citet{Bengaly2018}, with the addition of the \citet{Secrest2022} result.}
    \label{tab:DipoleRef}
\end{table*}

Figure \ref{fig:Dipole} shows the results of detecting the cosmic dipole for ET-CE and ET-2CE. It is clear that both ET-CE and ET-2CE are able to detect the cosmic dipole with the reported amplitude $\mathcal{D}\ga 0.01$ with high significance (Table \ref{tab:pValueAbs}). Even if $\mathcal{D}$ is as low as $0.005$, the detection significance will still reach $4\sigma$. 

\begin{table}
    \centering
    \begin{tabular}{c|ccc}
    \diagbox{network}{amplitude} & 0.005 & 0.01 & 0.02  \\ \hline
    ET-CE                                 & 4.27  & 6.21 & 12.13 \\ \hline
    ET-2CE                                & 3.89  & 8.64 & 13.41
    \end{tabular}
    \caption{The detection significance of cosmic dipole through the $a_{11}$ measurement,  for various the input dipole amplitudes.}
    \label{tab:pValueAbs}
\end{table}

\section{Conclusion} \label{sec:conclusion} 
In the era of precision cosmology, fundamental principles such as the cosmological principle will be tested stringently. One issue today is the inconsistency between CMB dipole and dipole of radio galaxy distribution.  GW surveys have unique advantage on probing the large scale cosmic anisotropies and possible violations of the cosmological principle, since GWs are transparent to the Galactic and extragalactic foregrounds.

In this work, we investigate the instrumental effects of GW detector networks and its impact on detecting cosmic anisotropy. We find that the $m=0$ spherical harmonic modes are heavily contaminated by the detector induced anisotropy at $\ell\la 10$. This anisotropy is correlated with the sky distribution of SNR and localization accuracy. Upgrading ET-CE to ET-2CE alleviates the problem, but does not solve it. Fortunately, the $m\neq 0$ modes are free of such issue, since rotation of the Earth averages out the instrumental anisotropy in longitude direction. This effect opens a clean window of detecting cosmic anisotropy in all $m\neq 0$ components. Through MC simulations, we show that the cosmic dipole reported in the literature with amplitude $\mathcal{D}\ga 0.01$ can be detected with high significance by both ET-CE and ET-2CE. 

The cosmic anisotropy may have appearances other than the cosmic dipole \citep{2022NewAR..9501659P,2023CQGra..40i4001K}. Furthermore, it must also have structures both across the sky and along the line of sight. All these can be probed by the 3D large scale distribution of GW sources in principle. The detector induced 3D clustering pattern and the existence of clean windows for cosmological inhomogeneity  detection will be further studied. This will further reveal  the capability of the GW based luminosity distance space large scale structure \citep{2018arXiv181011915Z,2021JCAP...01..036N,2021JCAP...02..035L,2021PhRvD.103j3507P,2022JCAP...02..003L}.  

\section*{Acknowledgements}
This work is supported by National Science Foundation of China (11621303 and 12273035), the National Key R\&D Program of China (2020YFC2201602, 2022YFC2200100, 2021YFC2203102) and CMS-CSST-2021-A02.

This work made use of the Gravity Supercomputer at the Department of Astronomy, Shanghai Jiao Tong University.

\section*{Data Availability}

We develop {\tt PYTHON} package {\tt GWSKY} for this work, the source code is available on GitHub \faGithub~ \url{https://github.com/mzLi01/gwsky}.



\bibliographystyle{mnras}
\bibliography{main} 








\bsp	
\label{lastpage}
\end{document}